\documentclass[onecolumn,superscriptaddress,preprintnumbers,amsmath,amssymb,prc,nofootinbib]{revtex4}

\usepackage{graphicx}
\usepackage{subfigure}
\usepackage{dcolumn}
\usepackage{bm}

\usepackage{color}

\begin{document}
\title{Electromagnetic fields in small systems from a multiphase transport model}

\author{Xin-Li Zhao}
\affiliation{Shanghai Institute of Applied Physics, Chinese Academy of Sciences, Shanghai 201800, China}
\affiliation{University of Chinese Academy of Sciences, Beijing 100049, China}
\author{Yu-Gang Ma}
\email[]{ygma@sinap.ac.cn}
\affiliation{Shanghai Institute of Applied Physics, Chinese Academy of Sciences, Shanghai 201800, China}
\author{Guo-Liang Ma}
\email[]{glma@sinap.ac.cn}
\affiliation{Shanghai Institute of Applied Physics, Chinese Academy of Sciences, Shanghai 201800, China}


\begin{abstract}

We calculate the electromagnetic fields generated in small systems by using a multiphase transport (AMPT) model. Compared to $A+A$ collisions, we find that the absolute electric and magnetic fields are not small in $p$+Au and $d$+Au collisions at energies available at the BNL Relativistic Heavy Ion Collider and in $p$+Pb collisions at energies available at the CERN Large Hadron Collider. We study the centrality dependencies and the spatial distributions of electromagnetic fields. We further investigate the azimuthal fluctuations of the magnetic field and its correlation with the fluctuating geometry using event-by-event simulations. We find that the azimuthal correlation $\left\langle \cos2(\Psi_B - \Psi_{2}) \right\rangle$ between the magnetic field direction and the second harmonic participant plane is almost zero in small systems with high multiplicities, but not in those with low multiplicities. This indicates that the charge azimuthal correlation, $\left\langle \cos(\phi_{\alpha}+\phi_{\beta} - 2\Psi_{RP}) \right\rangle$,  is not a valid probe to study the chiral magnetic effect (CME) in small systems with high multiplicities. However, we suggest searching for possible CME effects in small systems with low multiplicities.

\end{abstract}


\maketitle

\section{Introduction}
\label{introduction}

Relativistic heavy-ion collisions are believed to create a deconfined matter with partonic degrees of freedom, i.e. quark-gluon plasma (QGP), with increasing temperature or baryon chemical potential~\cite{Adams:2005dq,Adcox:2004mh,Song:2017wtw}. On the other hand, the relativistic motion of the colliding heavy ions also brings an extremely large electromagnetic field with a magnitude of about $eB \sim m_\pi^2 \sim 10^{18}$ to  $10^{19}$ Gauss from the energy available at the BNL Relativistic Heavy Ion Collider (RHIC energy) to the energy available at the CERN Large Hadron Collider (LHC energy)~\cite{Kharzeev:2007jp,Skokov:2009qp,Asakawa:2010bu,Voronyuk:2011jd,Ou:2011fm,Bzdak:2011yy,Deng:2012pc}. In the past few years,  strong interest has been generated, which has led to many great efforts to study various possible physical effects related to the strong electromagnetic fields. A very famous example is the so-called chiral magnetic effect (CME)~\cite{Kharzeev:2007jp,Kharzeev:2004ey,Kharzeev:2007tn,Fukushima:2008xe,Hattori:2016emy} in which the joint action from a nonzero axial charge density of the matter and the magnetic field $\textbf{B}$ will lead to a dipole charge separation along the $\textbf{B}$ direction. The charge azimuthal correlation or three-particle correlator ($\gamma=\left\langle \cos(\phi_{\alpha}+\phi_{\beta} - 2\Psi_{RP}) \right\rangle$=$\left\langle \cos(\phi_{\alpha}+\phi_{\beta} - 2\phi_{c}) \right\rangle$/$v_{2,c}$) was proposed to experimentally measure the CME~\cite{Voloshin:2004vk}. The first measurements are consistent with the CME expectations both in Au+Au collisions at RHIC energy~\cite{Abelev:2009ac,Abelev:2009ad} and Pb+Pb collisions at the LHC energy~\cite{Abelev:2012pa}, though many background effects can also contribute to the observable $\gamma$~\cite{Schlichting:2010qia,Pratt:2010zn,Bzdak:2009fc,Bzdak:2010fd,Liao:2010nv,Wang:2009kd,Bzdak:2012ia}. However, the recent measurements on small systems from the CMS experiment show similar magnitudes of charge azimuthal correlation in $p$+Pb collisions to those in Pb+Pb collisions at the LHC energy~\cite{Khachatryan:2016got,Sirunyan:2017quh}, which strongly challenges the CME interpretation due to the following arguments. For $A+A$ collisions, the major axis of eccentricity is almost aligned closely with the magnetic field. It is the special type of configuration that makes the CME maximally measured by the three-particle correlator. However, for $p+A$ collisions, the major axis of the eccentricity is found to be random with respect to the magnetic field~\cite{Belmont:2016oqp}. This indicates that the magnetic field direction and the participant plane are almost irrelevant in $p+A$ collisions. As a result, the `relevance vs irrelevance' between $A+A$ and $p+A$ implies that the correlator due to the CME should be very different between large and small systems. Therefore, the observed similar magnitude and multiplicity dependence of the three-particle correlator indicate that the dominant contribution of the correlation signal may not be related to the CME~\cite{Khachatryan:2016got,Sirunyan:2017quh,Belmont:2016oqp}.

The aim of this work is to give a detailed study of the space structure of event-by-event-ly generated electromagnetic fields in small systems, e.g. $p$+Au, $d$+Au, $p$+Pb collisions. Because one expects that the charge separation signal $\Delta\gamma$ is proportional to $\left\langle B^{2}\cos2(\Psi_B - \Psi_{EP}) \right\rangle$~\cite{Bloczynski:2012en}, we focus on both the magnitude of the magnetic field $\textbf{B}$ and its azimuthal correlation $\left\langle \cos2(\Psi_B - \Psi_{2}) \right\rangle$ with the bulk matter geometry  (represented by the second order of event plane $\Psi_{2}$). We will show the absolute electric and magnetic fields are not small in $p$+Au and $d$+Au collisions at the RHIC energy and in $p$+Pb collisions at the LHC energy. On the other hand, because $\Psi_B$ and $\Psi_{2}$ can fluctuate due to the fluctuations of spectator~\cite{Bzdak:2011yy,Deng:2012pc} and participant nucleons~\cite{Alver:2010gr,Ma:2010dv, Bhalerao:2011yg,Teaney:2010vd,Qiu:2011iv,Staig:2010pn}, a finite $\left\langle \cos2(\Psi_B - \Psi_{2}) \right\rangle$ is the key to experimentally measure any magnetic field induced effect through the observable $\gamma$.  It has been turned out that for both the $\textbf{B}$ direction and the geometry plane (event plane or participant plane) are strongly correlated in $A+A$ collisions~\cite{Bloczynski:2012en}, but are not correlated in $p+A$ collisions~\cite{Belmont:2016oqp}. Nevertheless, we show their orientations are fully random or only weakly correlated in small systems with high multiplicities, but not in those with low multiplicities. We perform our calculations by using a multiphase transport model (AMPT) model~\cite{Lin:2004en}. The model is suitable to study both electromagnetic fields and the geometry fluctuations for two reasons. First, the HIJING model~\cite{Wang:1991hta,Gyulassy:1994ew}, which is the initial part of the AMPT model, has been successfully applied to calculate the properties of electromagnetic fields~\cite{Deng:2012pc}. Meanwhile, the AMPT model also can give very good descriptions to anisotropic flows originating from geometry fluctuations in both $A+A$ and $p+A$ systems at both RHIC energy and the LHC energy~\cite{Ma:2016fve,Adare:2015ctn,Bzdak:2014dia}. Therefore, we choose it as our tool to study the properties of electromagnetic fields in small systems in this work.

This paper is organized as follows. We give a general setup of our calculations in Sec. \ref{GS}. In Sec. \ref{results}, we present our numerical results. We expand discussions and summarize in Sec. \ref{summary}.
\section{GENERAL SETUP}
\label{GS}
\subsection{AMPT model}
The AMPT model with a string-melting mechanism is a dynamical transport model~\cite{Lin:2004en}, which consists of four main components: initial condition, parton cascade, hadronization, and hadronic rescatterings. The initial condition, which includes the spatial and momentum distributions of participant matter, minijet partons production, and soft string excitations, is obtained through the HIJING model~\cite{Wang:1991hta,Gyulassy:1994ew}. The parton cascade starts the partonic evolution with a quark-antiquark plasma from the melting of strings. Parton scatterings are modeled by Zhang's parton cascade (ZPC)~\cite{Zhang:1997ej}, which currently only includes two-body elastic parton scatterings using cross sections from the pQCD with screening masses. A naive quark coalescence model is then used to combine partons into hadrons~\cite{Lin:2001zk} when the system freezes out. The evolution dynamics of the hadronic matter is described by a relativistic transport (ART) model~\cite{Li:1995pra}.  For $p$+Au, $d$+Au and $p$+Pb collisions, we set a random orientation of the reaction plane in our AMPT simulations, in order to mimic the experimental cases.

\subsection{Calculations of the electromagnetic field and the participant plane}
We calculate electromagnetic fields by using the spatial distribution of protons from the initial condition i.e. the HIJING model. The magnetic field is calculated specifically at the center-of-mass frame, in which ${\bf r}_c = (x_c$ , $y_c$ , $z_c)$ is the center of mass of the participating nucleons in small system collisions. In this calculation, following in Ref.~\cite{Deng:2012pc}, we calculate the electromagnetic fields as
\begin{eqnarray}
e{\bf E}(t,{\bf r})&=&\frac{e^2}{4\pi}{\sum\limits_{n}}Z_{n} \frac{{\bf R}_n- R_n{\bf v}_n}{(R_n-{\bf R}_n \cdot {\bf v}_n)^3}(1-v_n^2),
\label{elec}\\
e{\bf B}(t,{\bf r})&=&\frac{e^2}{4\pi}{\sum\limits_{n}}Z_{n} \frac{{\bf v}_n \times {\bf R}_n}{(R_n-{\bf R}_n \cdot {\bf v}_n)^3}(1-v_n^2).
\label{magn}
\end{eqnarray}%
Here we use natural unit  $\hbar = c = 1$, where $Z_n$ is the charge number of the nth particle, for proton it is one, ${\bf R}_n = {\bf r} - {\bf r}_n$ is the relative position of the field point {\bf r} to the source point ${\bf r}_n$, and ${\bf r}_n$ is the location of the nth particle with velocity ${\bf v}_n$ at the retarded time $t_{n} = t - |{\bf r} - {\bf r}_n|$. The summations run over all charged protons in the system. When we calculate the electromagnetic fields of the small systems, we consider all protons that are at least $r > 0.3$ fm~\cite{Bzdak:2011yy} away from the center of mass ${\bf r_c}$.

We calculate the second order of participant plane $\Psi_2$ by using the spatial distribution of partons from the string-melting mechanism before the parton cascade process starts. For $p$+Au, $d$+Au and $p$+Pb collisions, we calculate the angle of participant plane $\Psi_2$ by
\begin{eqnarray}
\Psi_{2}&=&\frac{\arctan2(\left\langle r^{2}\sin(2\phi) \right\rangle, \left\langle r^{2}\cos(2\phi) \right\rangle)+\pi}{2},
\label{psi2}
\end{eqnarray}%
where $r$ is the displacement of the participating partons from the center of mass, $\phi$ is the azimuthal angle of the participating partons in the transverse plane, and the brackets $\left\langle  \right\rangle$ mean taking average over all participating partons~\cite{Alver:2010gr,Ma:2010dv}.
\section{Results and discussions}
\label{results}
\subsection{Centrality dependencies}
\label{resultsA}

\begin{figure}[htbp]
\begin{minipage}[t]{0.31\linewidth}
\centering
\includegraphics[width=1.05\textwidth]{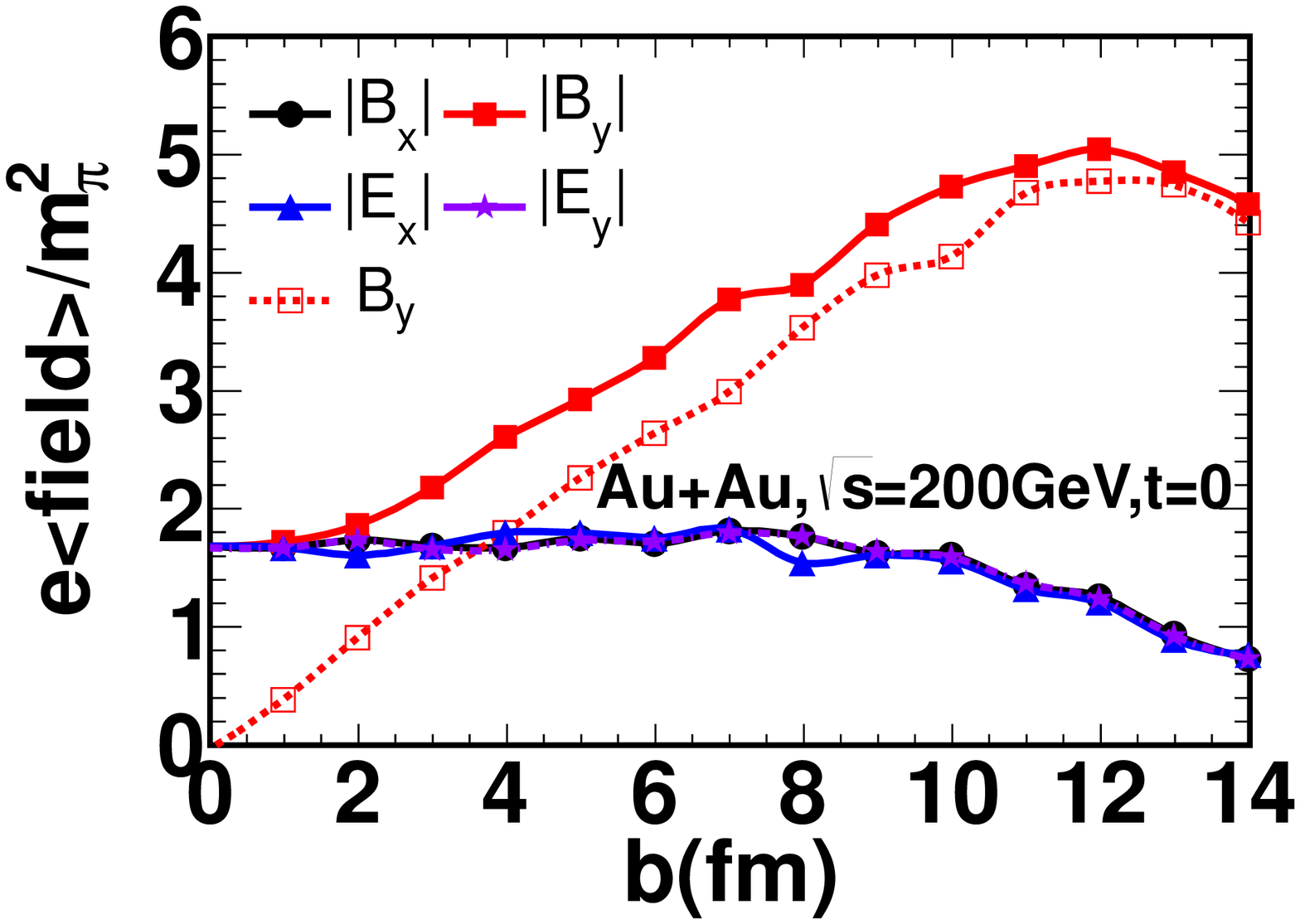}
\caption{(Color online) The electromagnetic fields at $t = 0$ and ${\bf r}={\bf 0}$ for Au+Au collisions at $\sqrt s = 200$ GeV.}\label{fig:AuAu}
\end{minipage}%
\hfill
\begin{minipage}[t]{0.31\linewidth}
\centering
\includegraphics[width=1.05\textwidth]{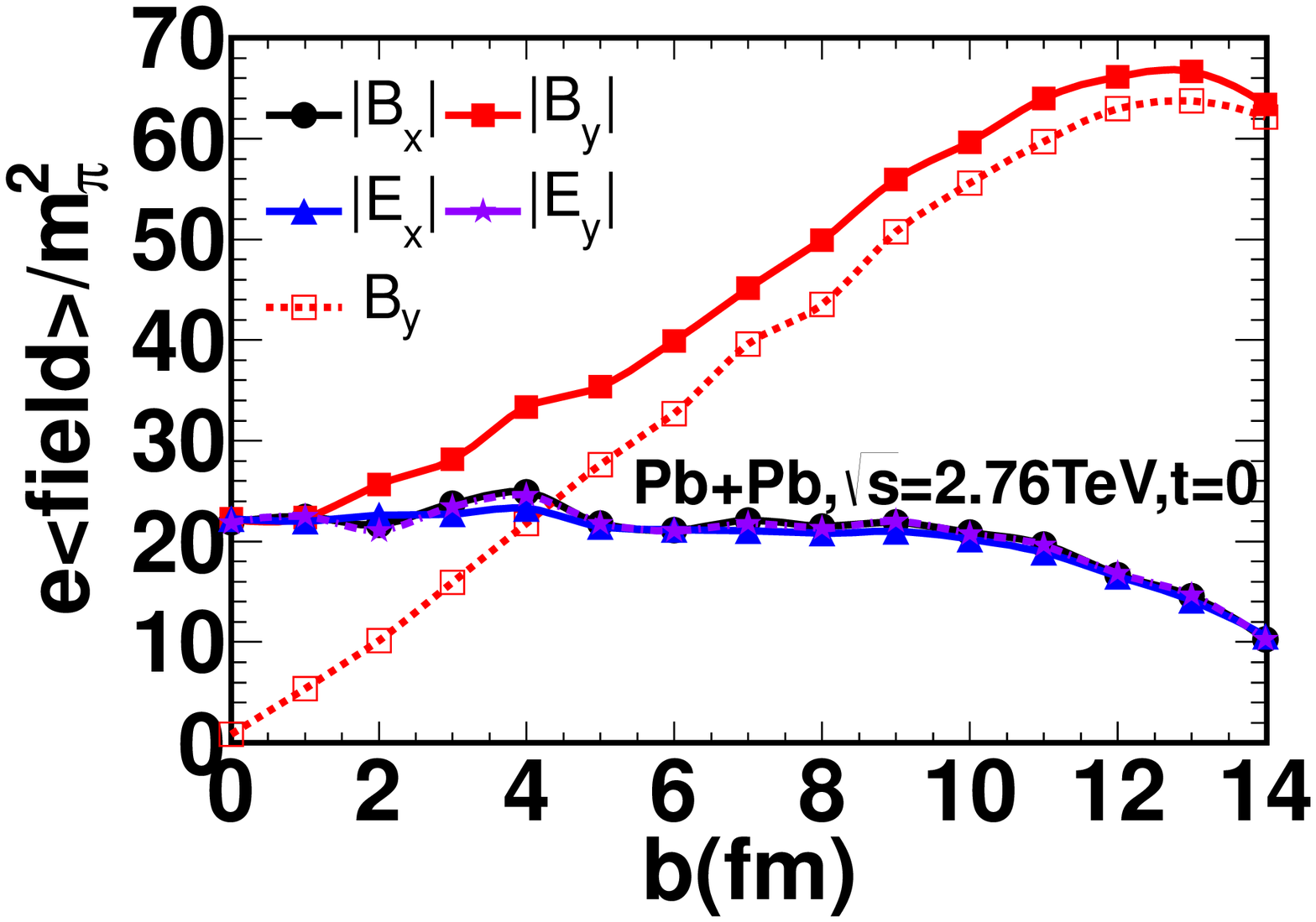}
\caption{(Color online) The electromagnetic fields at $t = 0$ and ${\bf r}={\bf 0}$ for Pb+Pb collisions at $\sqrt s = 2.76$ TeV.}\label{fig:PbPb}
\end{minipage}%
\hfill
\begin{minipage}[t]{0.31\linewidth}
\centering
\includegraphics[width=0.9\textwidth]{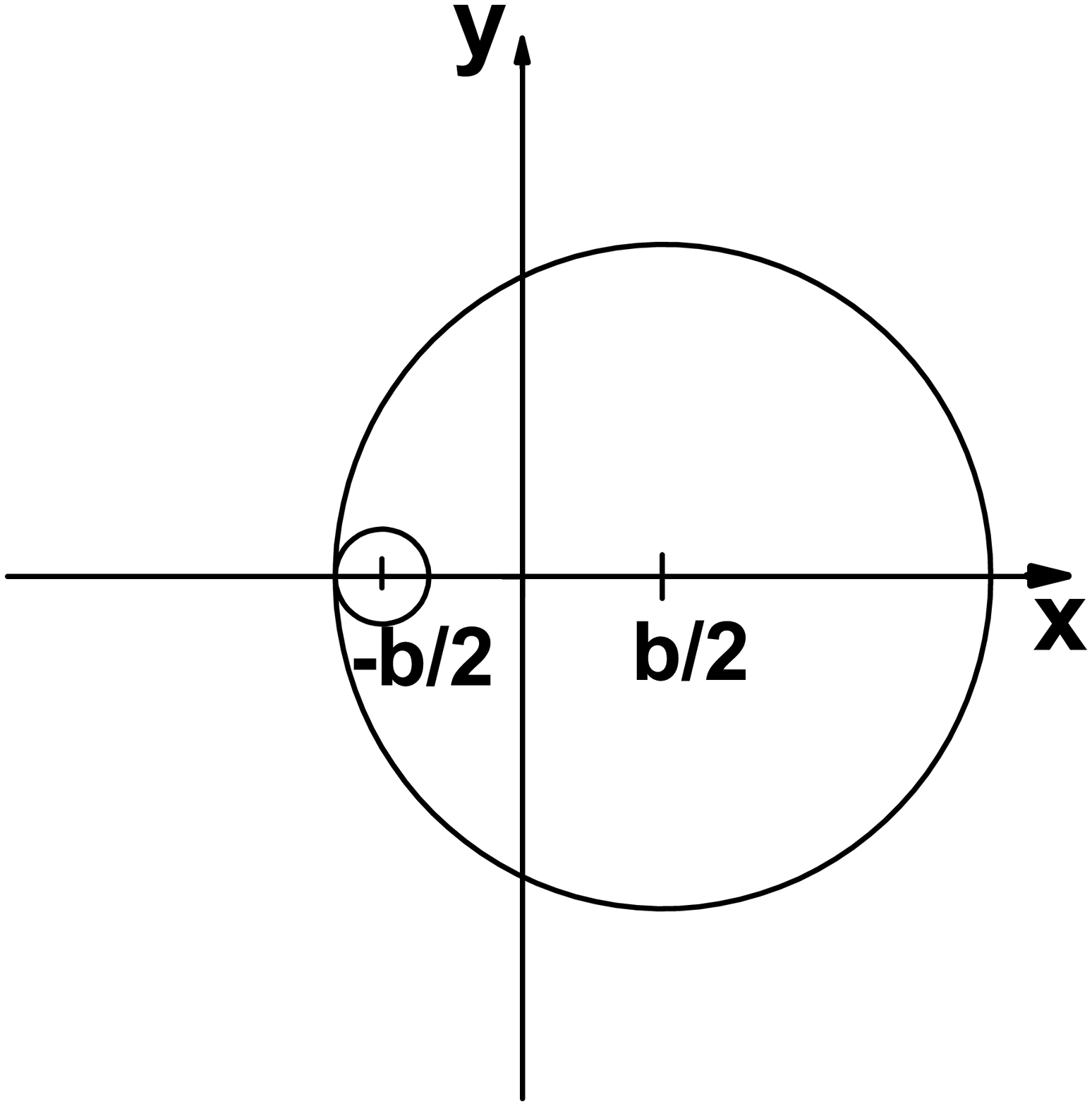}
\caption{The special geometrical illustration for $p$+Pb collisions with an impact parameter of $b$.}\label{fig:special}
\end{minipage}
\end{figure}

We first show the impact parameter $b$ dependencies of the electromagnetic fields at $t = 0$.  Fig.~\ref{fig:AuAu} shows the results for Au + Au collisions at $\textbf{r} = {\bf 0}$ and $t = 0$ at RHIC energy $\sqrt s = 200$ GeV, and Fig.~\ref{fig:PbPb} is for Pb+Pb collisions at $\textbf{r} = {\bf 0}$ and $t = 0$ at the LHC energy $\sqrt s = 2.76$ TeV (Note that the $\left\langle |B_x| \right\rangle$ and $\left\langle |E_{x,y}| \right\rangle$ look overlapped in the two figures.). Our calculations well reproduce the results of Deng and Huang~\cite{Deng:2012pc}.

\begin{figure*}[htb]
  \begin{minipage}[t]{0.333\linewidth}
    \includegraphics[width=0.95\textwidth]{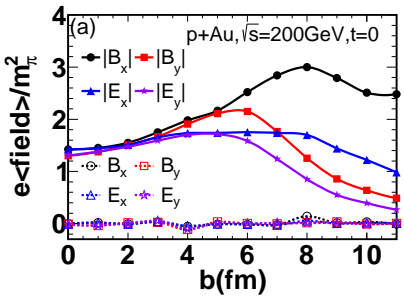}
    \label{fig:side:a}
  \end{minipage}%
  \begin{minipage}[t]{0.333\linewidth}
    \centering
    \includegraphics[width=0.95\textwidth]{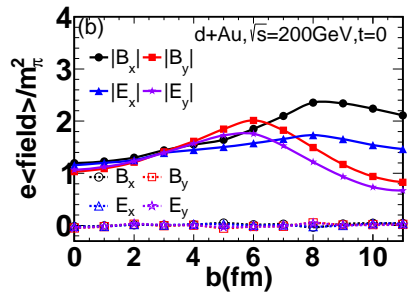}
    \label{fig:side:b}
  \end{minipage}%
  \begin{minipage}[t]{0.333\linewidth}
    \centering
    \includegraphics[width=0.95\textwidth]{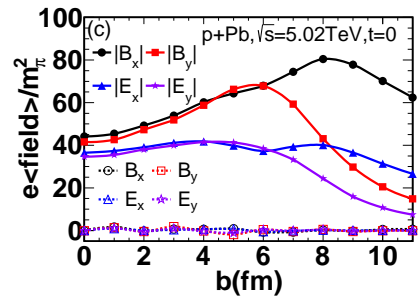}
    \label{fig:side:b}
  \end{minipage}
\caption{(Color online) The electromagnetic fields at $t = 0$ and ${\bf r}={\bf r}_c$ as functions of the impact parameter $b$ in $p$+Au collisions (a) and $d$+Au collisions (b) at $\sqrt s = 200$ GeV, and $p$+Pb collisions at $\sqrt s = 5.02$ TeV(c). } \label{fig:bdep}
\end{figure*}

Then we calculate the electromagnetic fields in three small systems, $p$+Au collisions and $d$+Au collisions at the RHIC energy $\sqrt s = 200$ GeV, and $p$+Pb collisions at the LHC energy $\sqrt s = 5.02$ TeV. Fig.~\ref{fig:special} illustrates a special configuration of $p$+Pb collision in which $x$ axis is along the impact parameter $b$ from proton (-$b/2$, 0 fm) to a Pb nucleus (0, $b/2$ fm) and $y$ axis is perpendicular to it. In general, for a given impact parameter $b$, the proton (Pb) center can be located around the circle with a radius of $b$ and the center at the Pb (proton) center. Because one experimentally can measure the event plane direction only according to the momentum anisotropies of the final particles, we define that the $x$ direction is along the direction of the participant plane $\Psi_2$ and the $y$ direction is perpendicular to $\Psi_2$. Their impact parameter dependencies of electromagnetic fields are shown in Fig.~\ref{fig:bdep}.  We can see that the three small systems show a similar behaviour in terms of impact parameter dependence of electromagnetic fields that shows $\left\langle |B_x| \right\rangle, \left\langle |B_y| \right\rangle, \left\langle |E_x| \right\rangle, \left\langle |E_y| \right\rangle$ first increase and then decrease. $\left\langle |B_y| \right\rangle$ and $\left\langle |E_y| \right\rangle$ have a maximum value at impact parameter $b \approx R_{A}$ and $R_{A}$ is the radius of the Au or Pb nucleus. $\left\langle |B_x| \right\rangle$ and $\left\langle |E_x| \right\rangle$ begin to decrease at impact parameter $b \approx 8$ fm. However, $\left\langle B_x \right\rangle \approx \left\langle B_y \right\rangle \approx \left\langle E_x \right\rangle \approx \left\langle E_y \right\rangle \approx 0$, that means the event-averaged electromagnetic fields are almost zero in the three small systems. Note that even $\left\langle B_y \right\rangle$ is zero in small systems but which is finite in $A+A$ collisions, because $\Psi_2$ is not correlated with the impact parameter in small systems.

\begin{figure*}[htb]
  \begin{minipage}[t]{0.333\linewidth}
    \includegraphics[width=0.94\textwidth]{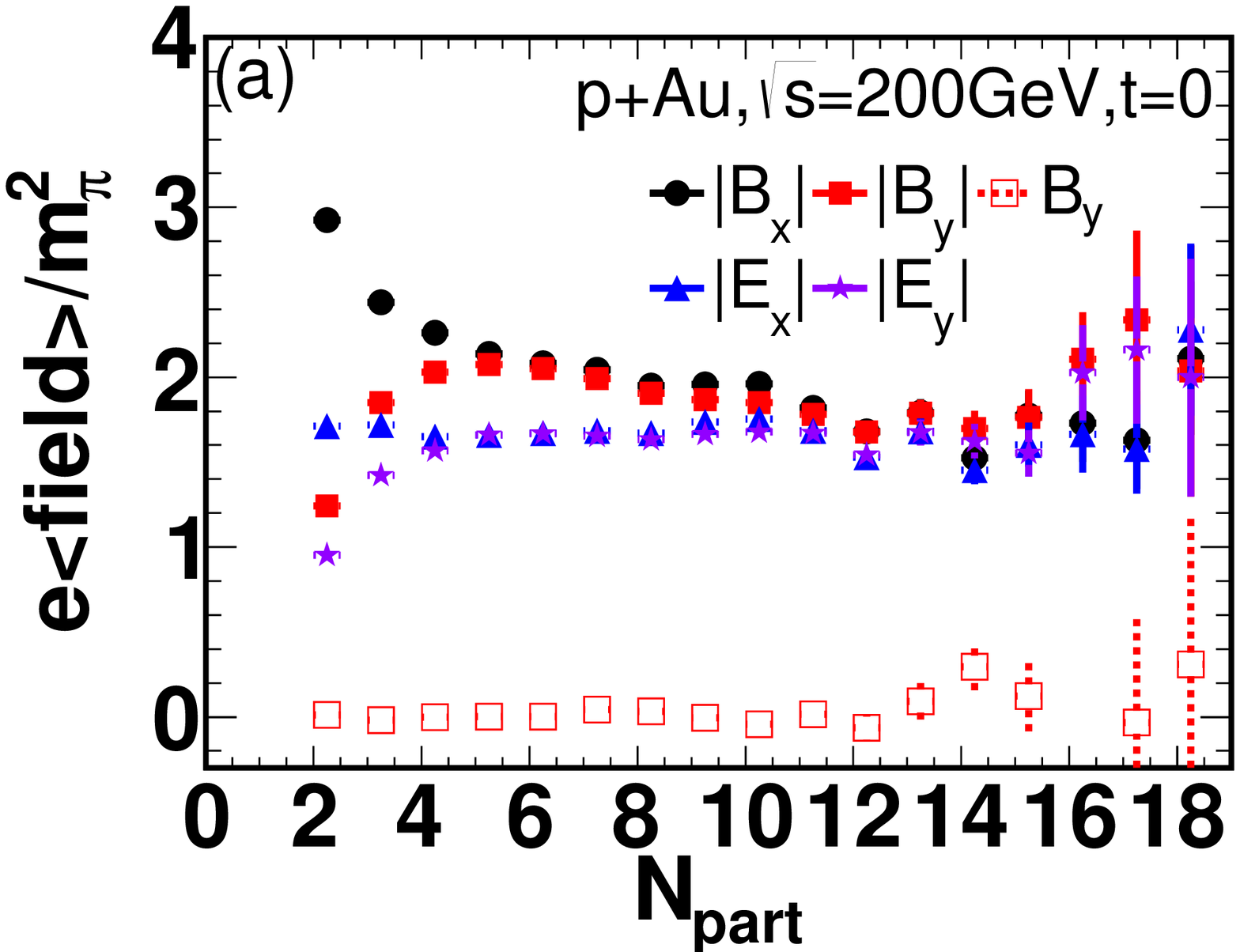}
    \label{fig:side:a}
  \end{minipage}%
  \begin{minipage}[t]{0.333\linewidth}
    \centering
    \includegraphics[width=0.94\textwidth]{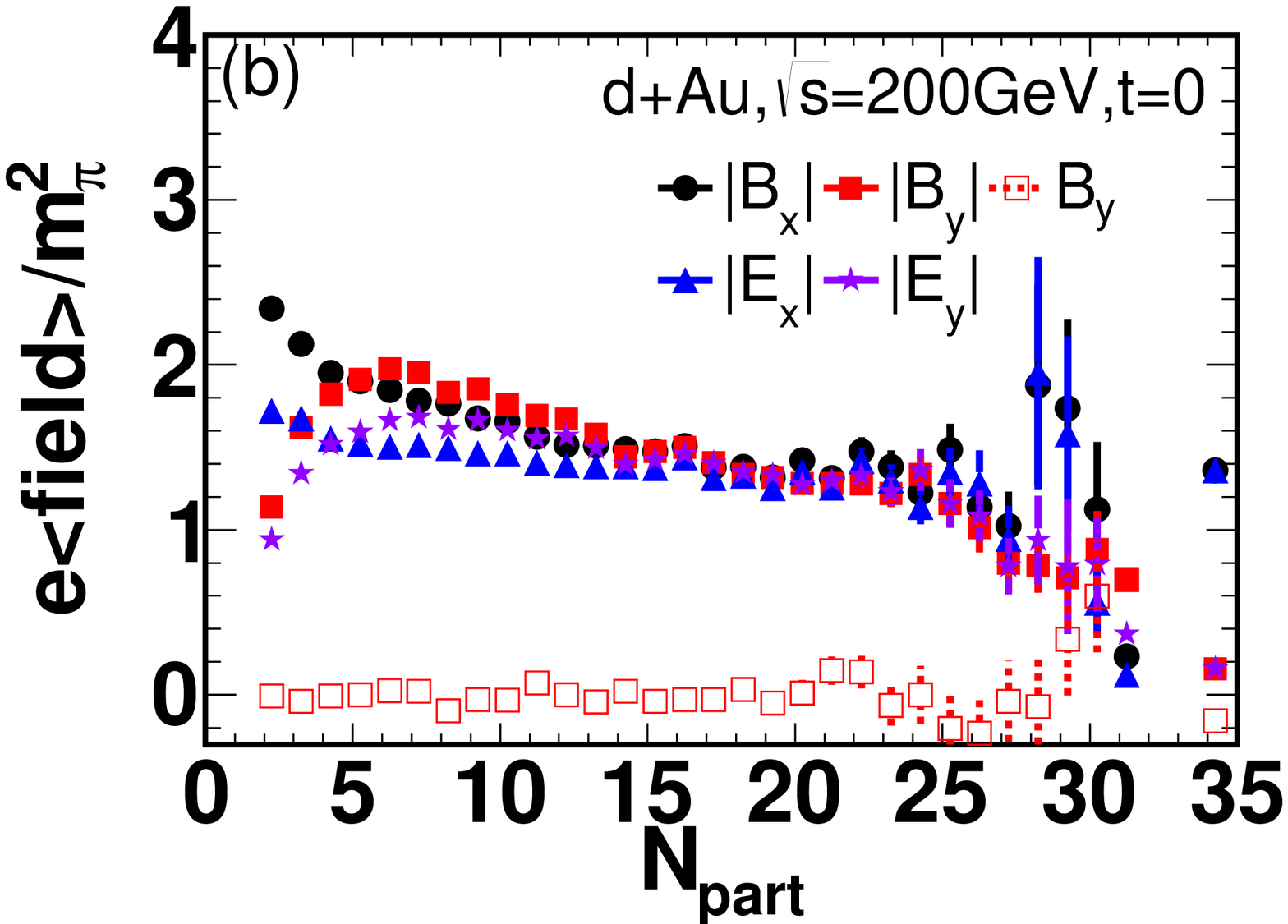}
    \label{fig:side:b}
  \end{minipage}%
  \begin{minipage}[t]{0.333\linewidth}
    \centering
    \includegraphics[width=0.94\textwidth]{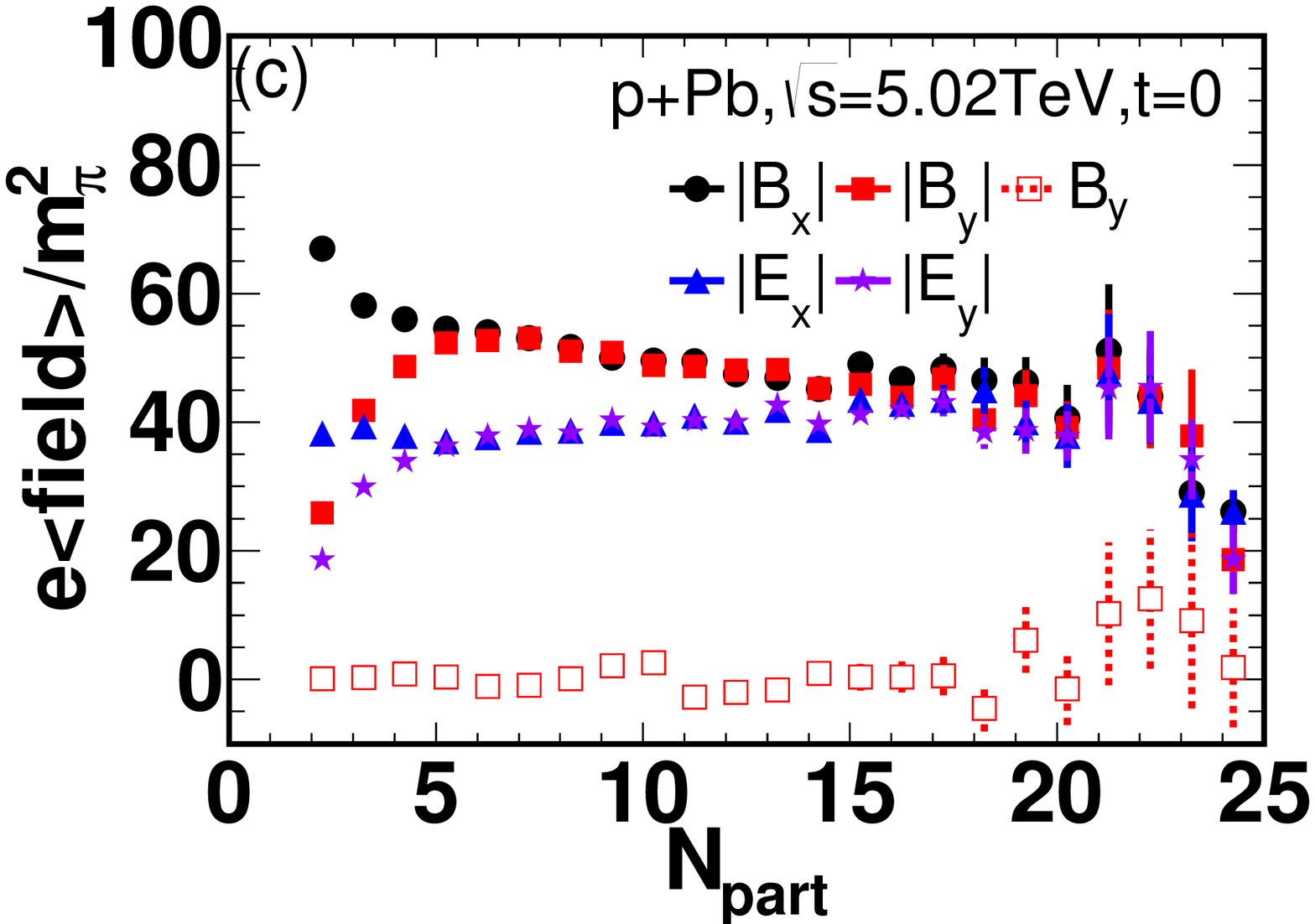}
    \label{fig:side:b}
  \end{minipage}
\caption{(Color online) The electromagnetic fields at $t = 0$ and ${\bf r}={\bf r}_c$ as functions of $N_{\mathrm{part}}$ in $p$+Au collisions (a) and $d$+Au collisions (b) at $\sqrt s = 200$ GeV, and $p$+Pb collisions at $\sqrt s = 5.02$ TeV(c).} \label{fig:Npart}
\end{figure*}

\begin{figure*}[htb]
  \begin{minipage}[t]{0.333\linewidth}
    \includegraphics[width=0.94\textwidth]{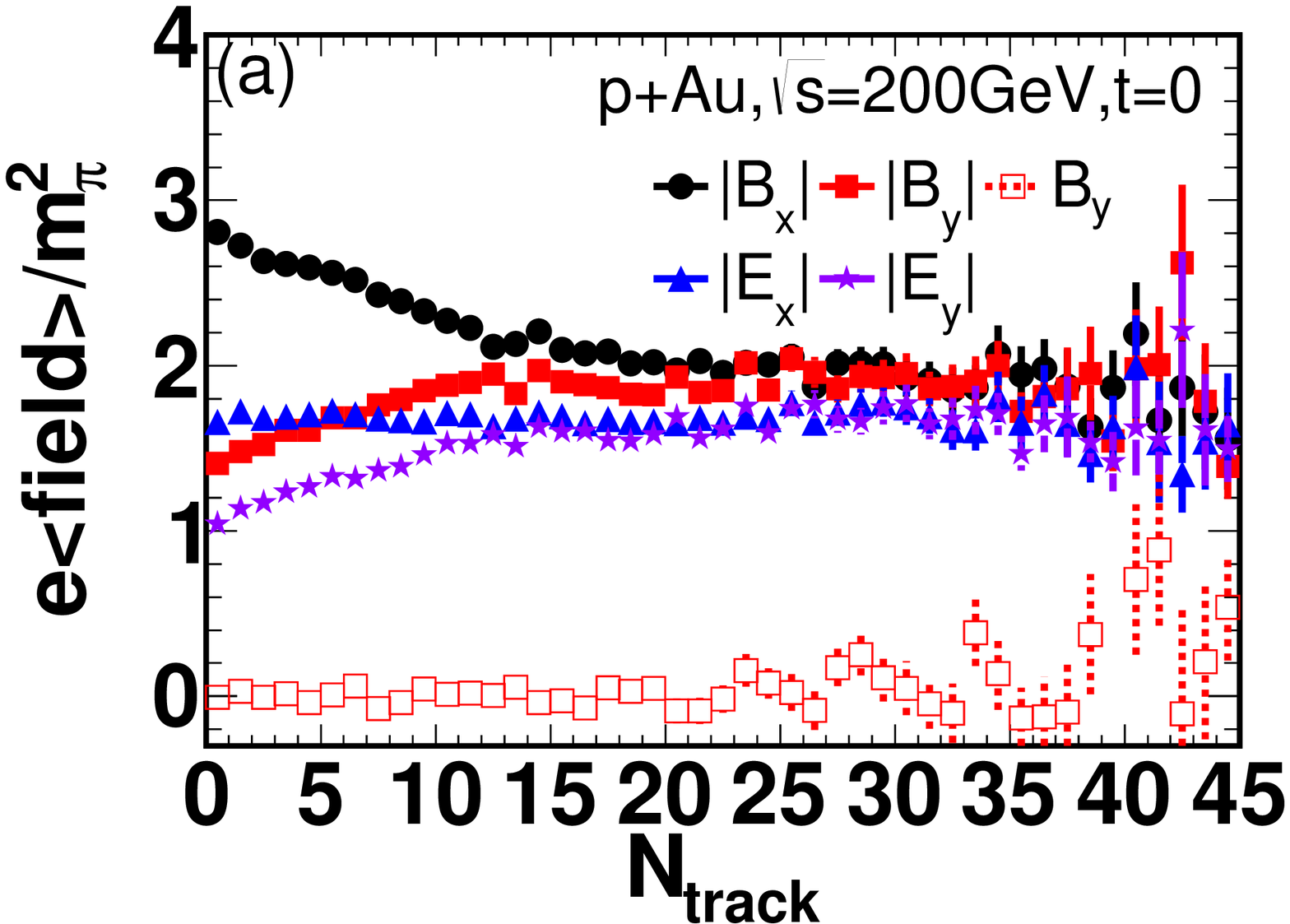}
    \label{fig:side:a}
  \end{minipage}%
  \begin{minipage}[t]{0.333\linewidth}
    \centering
    \includegraphics[width=0.94\textwidth]{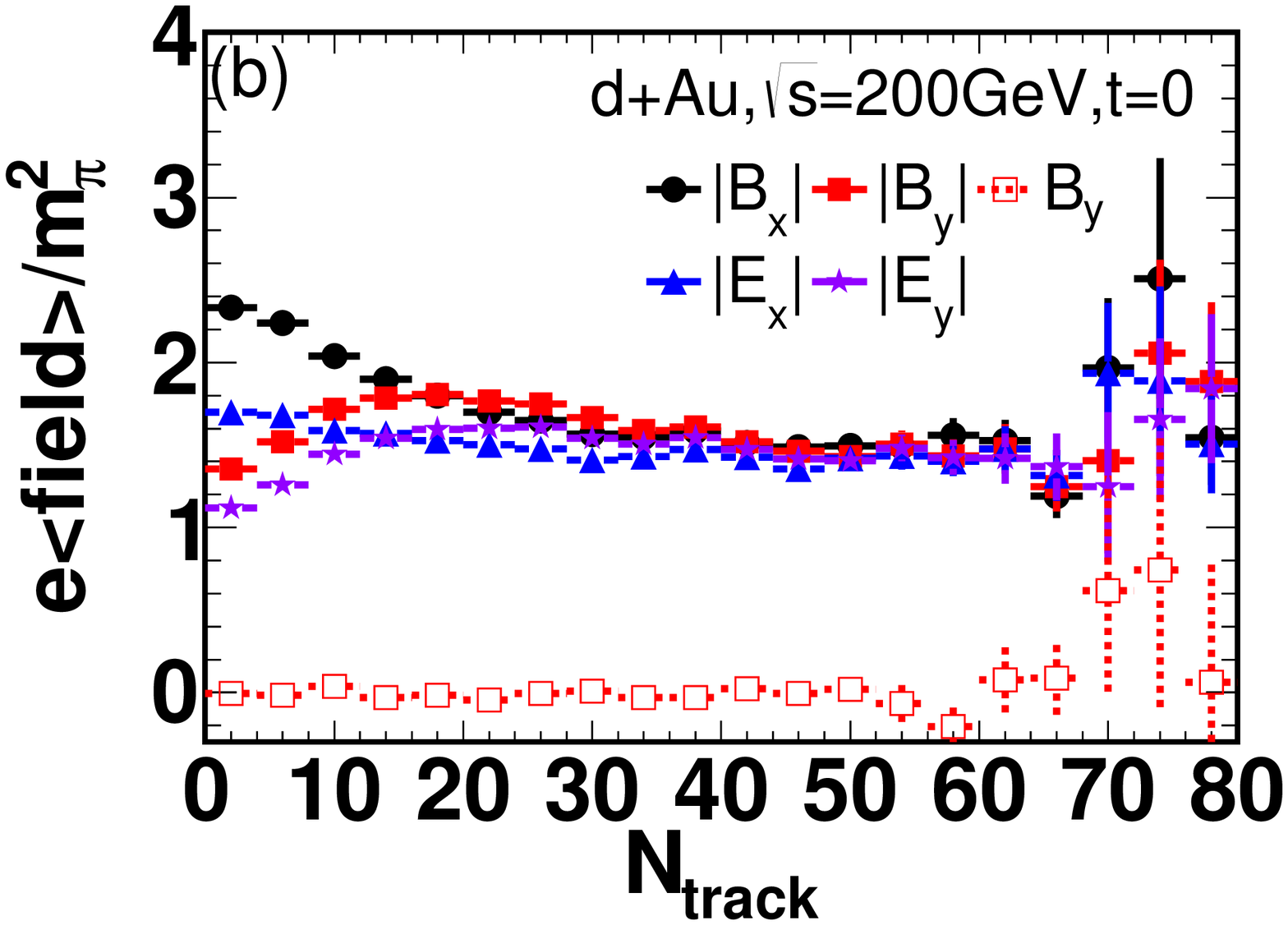}
    \label{fig:side:b}
  \end{minipage}%
  \begin{minipage}[t]{0.333\linewidth}
    \centering
    \includegraphics[width=0.94\textwidth]{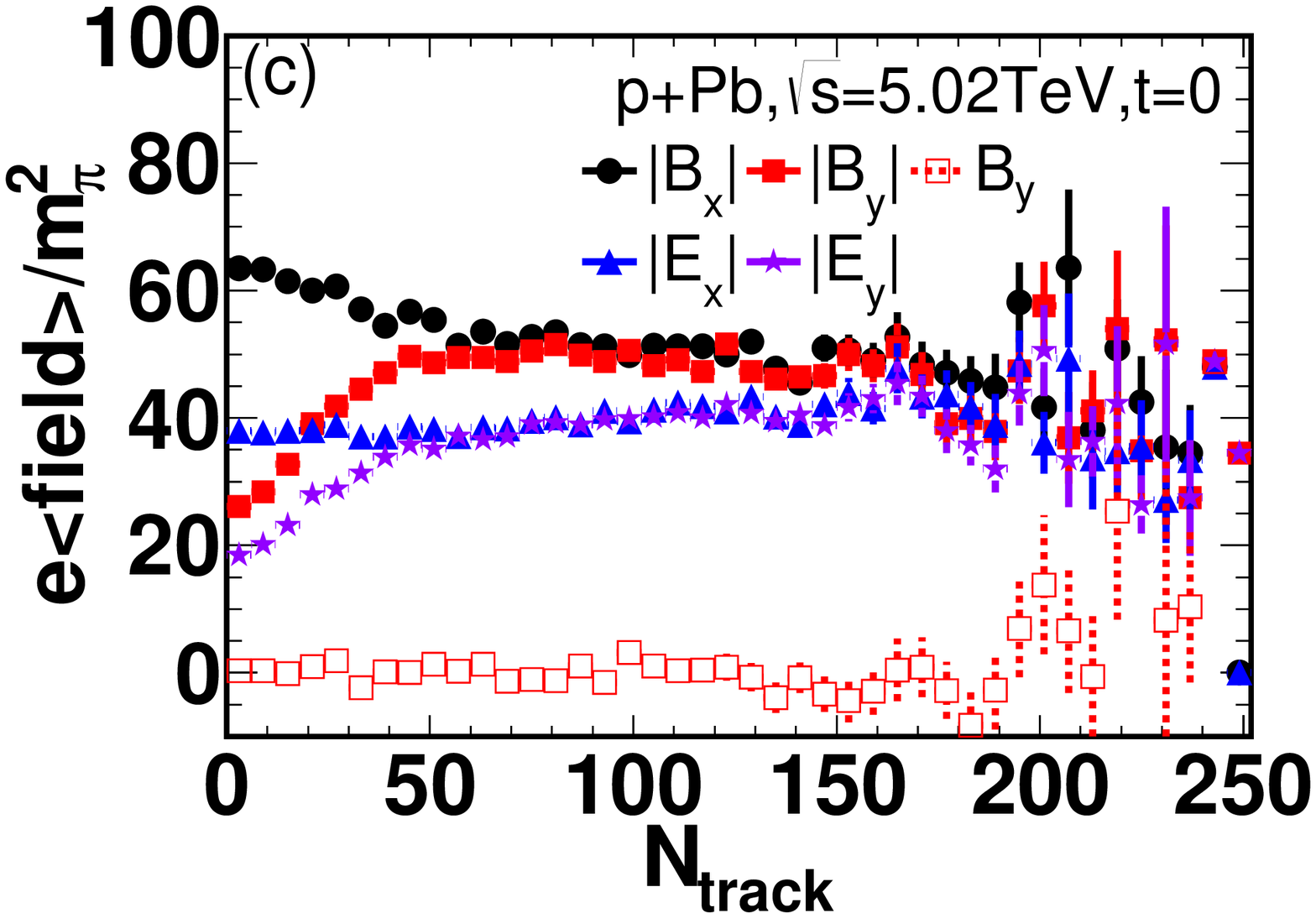}
    \label{fig:side:b}
  \end{minipage}
\caption{(Color online) The electromagnetic fields at $t = 0$ and ${\bf r}={\bf r}_c$ as functions of $N_{\mathrm{track}}$ in $p$+Au collisions (a) and $d$+Au collisions (b) at $\sqrt s = 200$ GeV, and $p$+Pb collisions at $\sqrt s = 5.02$ TeV(c).} \label{fig:Ntrack}
\end{figure*}

We show $N_{\mathrm{part}}$ (the number of participant nucleons) dependence of the electromagnetic fields at the center of mass ${\bf r}_c$ and $t = 0$ in Fig.~\ref{fig:Npart}. We can find that our $N_{\mathrm{part}}$ dependencies of magnetic field are consistent with the results from Ref.~\cite{Belmont:2016oqp}. In Fig.~\ref{fig:Ntrack}, we show the $N_{\mathrm{track}}$ dependencies of the electromagnetic fields at ${\bf r_c}$ and $t = 0$, where $N_{\mathrm{track}}$ is the number of final charged particles, which is an important experimental variable~\cite{Khachatryan:2016got}. Here we set $|\eta|<2.4$ and $p_T >0.4$ GeV/c at the LHC energy and $|\eta|<1$ and $p_T >0.15$ GeV/c at RHIC energy to match the CMS definition or the STAR acceptance of $N_{\mathrm{track}}$. We can see that although the averaged electromagnetic fields are zero (Only $\left\langle B_y \right\rangle$ is shown for clarification here), the average of the absolute value of the magnetic fields is still strong over the whole $N_{\mathrm{track}}$ range, which could potentially bring some electromagnetic field based effects.

\subsection{Spatial distributions of electromagnetic fields}
\label{resultsB}

\begin{figure*}[htb]
  \setlength{\abovecaptionskip}{0pt}
  \setlength{\belowcaptionskip}{8pt}\centerline{\includegraphics[scale=0.95]{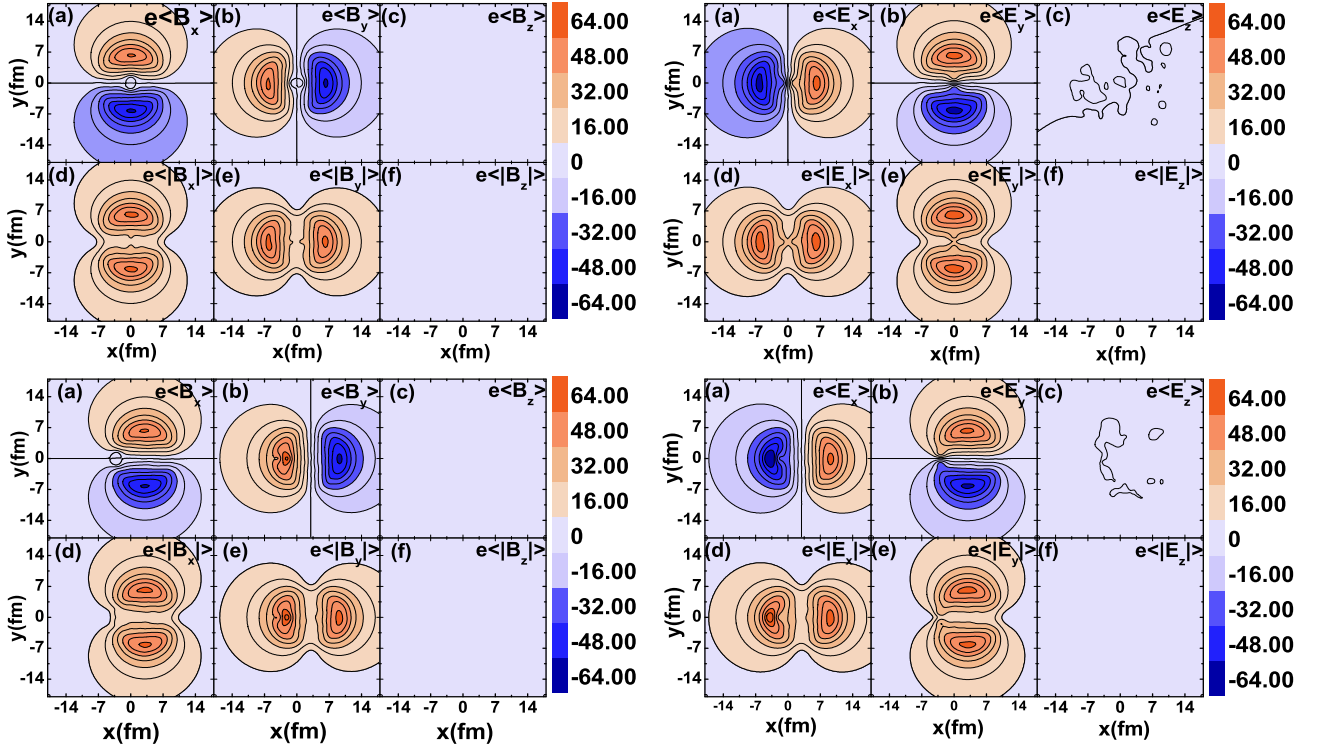}}
\caption{(Color online) The spatial distributions of the electromagnetic fields in the transverse plane at $t = 0$ for $b = 0$ (upper panels) and $b =$ 6 fm (lower panels) for $p$+Pb collisions at $\sqrt s = 5.02$ TeV, where the unit is $m_\pi^2$.}\label{fig:spa}
\end{figure*}

The spatial distributions of the electromagnetic fields are evidently inhomogeneous in Au+Au collisions and Pb+Pb collisions~\cite{Deng:2012pc}. In Fig.~\ref{fig:spa}, we show the contour plots of $\left\langle B_{x, y, z} \right\rangle$, $\left\langle |B_{x, y, z}| \right\rangle$, $\left\langle E_{x, y, z} \right\rangle$, $\left\langle |E_{x, y, z}| \right\rangle$ at $t = 0$ in the transverse plane for $p$+Pb collisions at $\sqrt s = 5.02$ TeV, where the upper two panels are for $b = 0$ fm and the lower two panels are for $b = 6$ fm. For $b = 6$ fm, we present a special configuration of $p$+Pb events in which $x$ axis is along the impact parameter $b$ from proton (-3, 0 fm) to Pb nucleus (0, 3 fm) and $y$ axis is perpendicular to it, as illustrated in Fig.~\ref{fig:special}.  We can see that for $p$+Pb collisions, longitudinal fields $B_z$, $E_z$, $|B_z|$ and $|E_z|$, are much smaller than the transverse fields $B_x$, $|B_x|$, $E_y$, and $|E_y|$. $B_x$, $|B_x|$, $E_y$, and $|E_y|$ peak around $(x$ , $y)=(b/2$ , $\pm R_{Pb})$, while $B_y$, $|B_y|$, $E_x$, and $|E_x|$ peak around $(x$ , $y)=(\pm R_{Pb}+b/2$ , $y=0)$ where $R_{Pb}$ is the radius of the Pb nucleus. Compared to Au+Au collisions~\cite{Deng:2012pc}, the spatial distributions of the magnetic fields are very different in $p$+Au and $p$+Pb collisions, because these small systems are unsymmetrical. Moreover, we also study the spatial distributions of electromagnetic fields for $p$+Au collisions at $\sqrt s = 200$ GeV and find that their spatial distributions are similar to those in $p$+Pb collisions, however, their fields have around $200/5020 \approx 1/25$ times smaller magnitudes than those in $p$+Pb collisions everywhere.

\subsection{Azimuthal correlations between magnetic fields and matter geometry}
\label{resultsC}

\begin{figure*}[htb]
  \setlength{\abovecaptionskip}{0pt}
  \setlength{\belowcaptionskip}{8pt}\centerline{\includegraphics[scale=0.6]{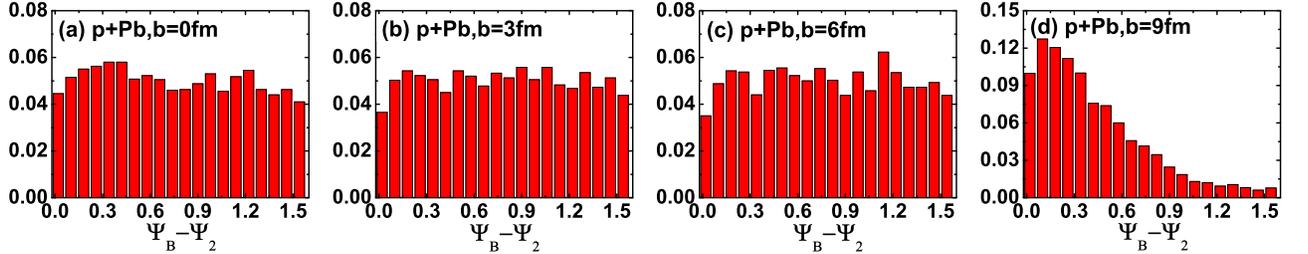}}
\caption{(Color online) The event-by-event histograms of $\Psi_{B}-\Psi_2$ at impact parameters $b =$ 0, 3, 6, 9 fm [(a)-(d)] for $p$+Pb collisions at $\sqrt s = 5.02$ TeV. Here $\Psi_{B}$ is the azimuthal direction of {\bf B} field (at $t = 0$ and ${\bf r}={\bf r}_c$) and $\Psi_2$ is the second harmonic participant plane.}\label{fig:psidiff}
\end{figure*}
\begin{figure*}[htb]
  \setlength{\abovecaptionskip}{0pt}
  \setlength{\belowcaptionskip}{8pt}\centerline{\includegraphics[scale=0.6]{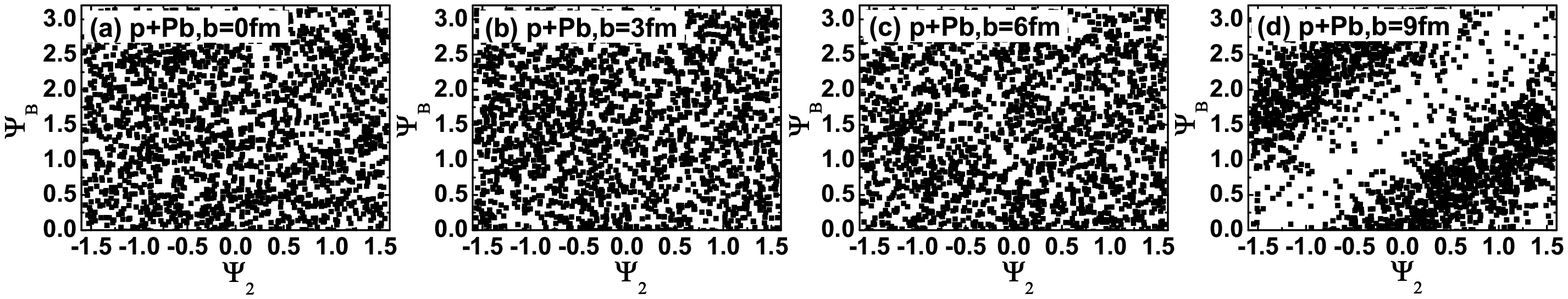}}
\caption{The scatter plots on $\Psi_{B}-\Psi_2$ plane at impact parameters $b =$ 0, 3, 6, 9 fm [(a)-(d)] for $p$+Pb collisions at $\sqrt s = 5.02$ TeV, where $\Psi_{B}$ is the azimuthal direction of {\bf B} field (at $t = 0$ and ${\bf r}={\bf r}_c$) and $\Psi_2$ is the second harmonic participant plane.}\label{fig:2Dplot}
\end{figure*}

As we mentioned, $\Delta\gamma$ is proportional to $\left\langle B^{2}\cos2(\Psi_B - \Psi_{2}) \right\rangle$. After showing that the magnitude of magnetic fields is not small, the correlation of $\left\langle \cos2(\Psi_B - \Psi_{2}) \right\rangle$ will be presented in this subsection. After we determine $\Psi_B$ and $\Psi_2$ event-by-event-ly, the distributions of their relative angle can be obtained. In Fig.~\ref{fig:psidiff} we plot the accumulated histograms of $\Psi_{B}-\Psi_2$ at $b =$ 0, 3, 6 and 9 fm for $p$+Pb collisions at $\sqrt s = 5.02$ TeV, where $\Psi_B$ is the azimuthal direction of the magnetic field at $t = 0$ and the center of mass of participants ${\bf r}_c$ and $\Psi_2$ is the second order of event plane calculated from Eq.~(\ref{psi2}).  For $b =$ 0, 3, and 6 fm, the histograms of $\Psi_{B}-\Psi_2$ are basically flat indicating that $\Psi_B$ and $\Psi_2$ are uncorrelated. For $b =$ 9 fm, the histogram has a shape peaking at $\Psi_{B}-\Psi_2 =$ 0 with a corresponding width. It is interesting to see that there is a correlation between $\Psi_B$ and $\Psi_2$ for very peripheral $p$+Pb collisions. This is because  when the impact parameter is very large, the projectile proton associated with a few wounded nucleons from the Pb side prefer to be in a dipole shape. Fig.~\ref{fig:2Dplot} shows the corresponding two-demisional correlation distributions. Again for $b =$ 0, 3 and 6 fm the events are almost uniformly distributed indicating negligible correlations between $\Psi_B$ and $\Psi_2$. For $b =$ 9 fm, the event distributions evidently concentrate around $\Psi_B - \Psi_2$ = 0 or $\pi$ indicating a nonzero correlation between the two angles.

\begin{figure}[htb]
\centering
\begin{minipage}[c]{0.4\textwidth}
\includegraphics[height=4.8cm,width=6.5cm]{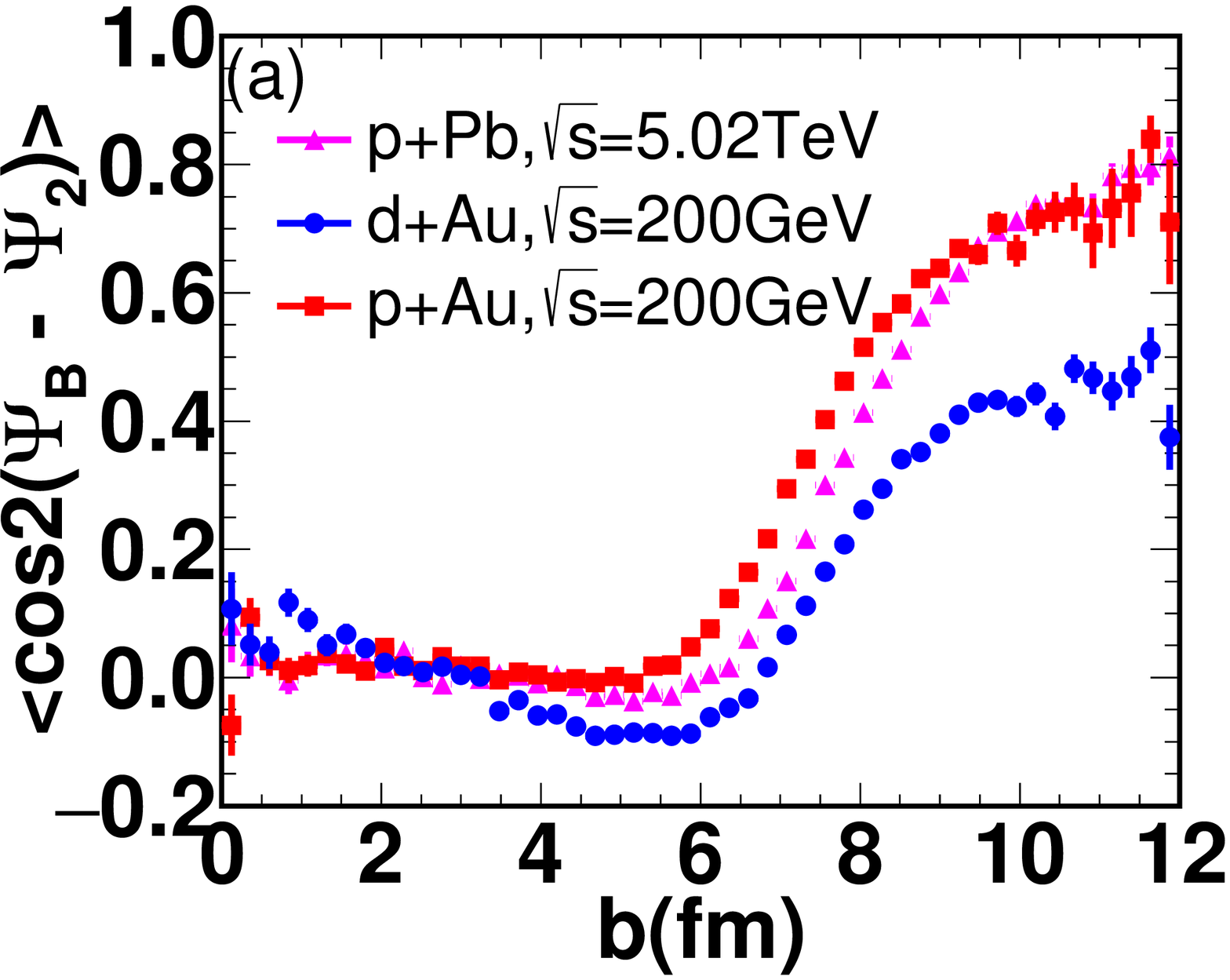}\end{minipage}
\begin{minipage}[c]{0.4\textwidth}
\includegraphics[height=4.8cm,width=6.5cm]{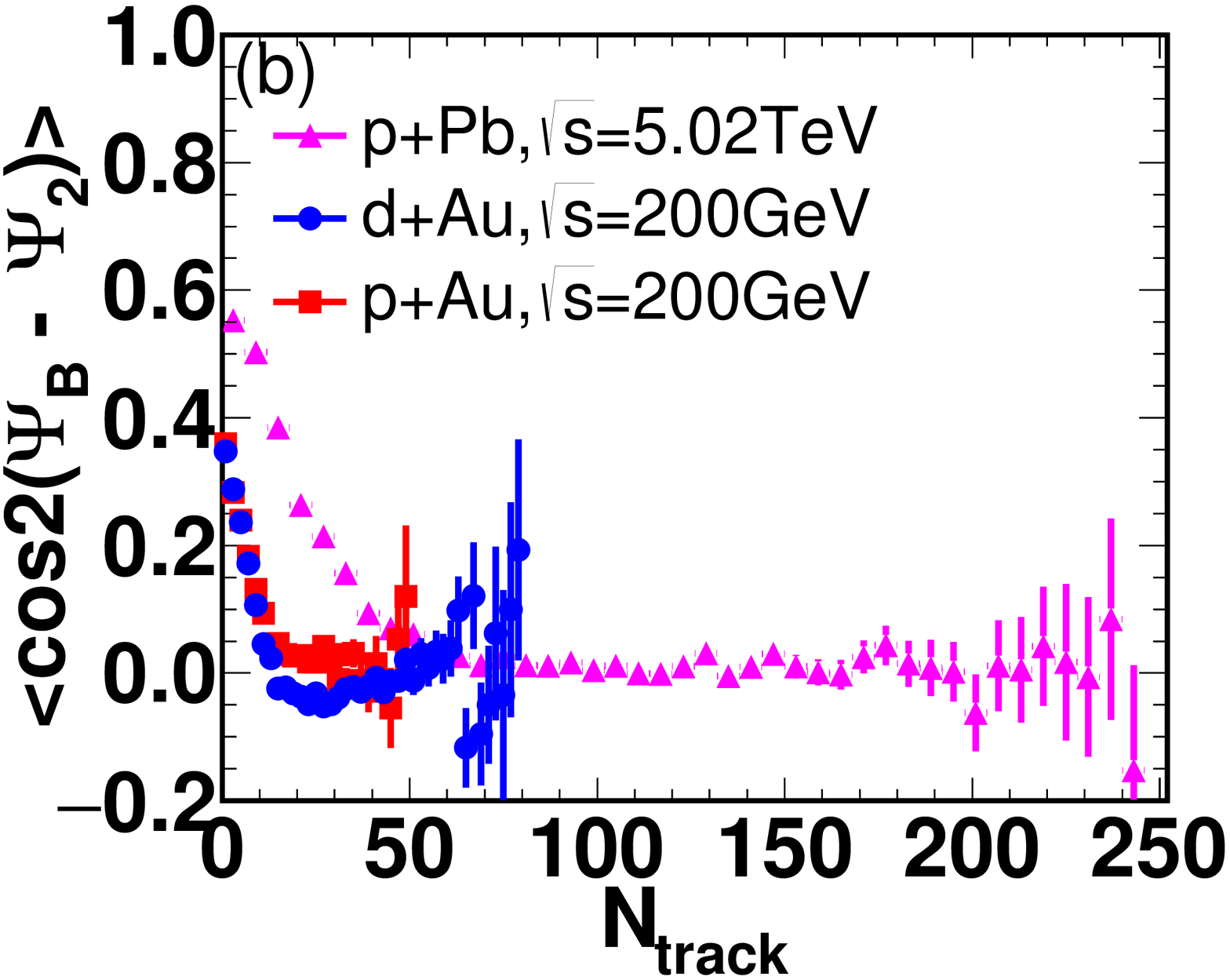}\end{minipage}
\caption{(Color online) The correlation $\left\langle \cos2(\Psi_B - \Psi_{2}) \right\rangle$ as functions of the impact parameter $b$ (a) and $N_{\mathrm{track}}$ (b) in $p$+Au and $d$+Au collisions at $\sqrt s = 200$ GeV, and $p$+Pb collisions at $\sqrt s = 5.02$ TeV.} \label{fig:corre}
\end{figure}

Fig.~\ref{fig:corre} (a) shows the correlation $\left\langle \cos2(\Psi_B - \Psi_{2}) \right\rangle$ as a function of the impact parameter $b$ and Fig.~\ref{fig:corre} (b) shows the correlation $\left\langle \cos2(\Psi_B - \Psi_{2}) \right\rangle$ as a function of $N_{\mathrm{track}}$, respectively. We can see that the correlation $\left\langle \cos2(\Psi_B - \Psi_{2}) \right\rangle \approx 0$ when impact parameter is not very big in three small systems $p$+Au, $d$+Au and $p$+Pb collisions. Meanwhile, for $p$+Au and $d$+Au collisions at the RHIC energy, the correlation $\left\langle \cos2(\Psi_B - \Psi_{2}) \right\rangle$ is almost zero when $N_{\mathrm{track}} > 10$. For $p$+Pb collisions at the LHC energy, the correlation $\left\langle \cos2(\Psi_B - \Psi_{2}) \right\rangle$ is close to zero when $N_{\mathrm{track}} > 50$. It indicates that the traditional CME observable $\Delta\gamma$ becomes not valid for studying the CME in small systems with high multiplicities, for which the CMS experiment measured in the region of $N_{\mathrm{track}} > 100$~\cite{Khachatryan:2016got,Sirunyan:2017quh}. Because the magnetic field direction is likely parallel to the event plane in small systems with low multiplicities, unlike $A+A$ collisions where $\Psi_{B}$ is perpendicular to $\Psi_{2}$, $\Delta\gamma$ should be reduced or even change sign in the presence of the CME for low-multiplicity events of small systems. Therefore, it will be very interesting to measure $\Delta\gamma$ in low-multiplicity (or peripheral) events to observe such an effect in the future experiments. \footnote{The recent results from the CMS event shape engineering measurement show a hint that the intercept parameter $b_{norm}$ is changed from zero to a negative value with the decreasing of multiplicity, which is qualitatively consistent with our expectation~\cite{Sirunyan:2017quh}.} However, we emphasize that our proposal to search for the CME in low-multiplicity events works under the assumption that the QGP is created in those events. 

\section{Conclusions}
\label{summary}
In summary, we have utilized the AMPT model to investigate the generation of the electromagnetic fields in small system $p$+Au, $d$+Au collisions at RHIC energy, and $p$+Pb collisions at the LHC energy.  Although after averaging over many events the electromagnetic fields are almost zero, the absolute values of the magnetic fields are nonzero on an event-by-event basis and they can still reach the order of several $m_\pi^2$. The spatial structure of the electromagnetic field is further studied and very inhomogeneous distributions are found. Because of large fluctuations of both the azimuthal orientation of the magnetic field and the participant plane in small system collisions, the azimuthal correlation $\left\langle \cos2(\Psi_B - \Psi_{2}) \right\rangle$ between the magnetic field direction and the second harmonic participant plane is strongly suppressed in high-multiplicity events, but not in low-multiplicity events. The correlation $\left\langle \cos2(\Psi_B - \Psi_{2}) \right\rangle$ is almost zero in small systems with high multiplicities, which indicates that the traditional CME observable $\Delta\gamma$ is not valid to study the CME in small systems with high multiplicities. However, because of a nonzero correlation of $\left\langle \cos2(\Psi_B - \Psi_{2}) \right\rangle$ in low-multiplicity small system collisions, we suggest measuring $\Delta\gamma$ in small systems with low-multiplicities to search for some possible effects from the CME.
\begin{acknowledgments}
We thank Wei-Tian Deng, Xian-Gai Deng, Xu-Guang Huang, Qi-Ye Shou, Zhoudunming Tu, and Fuqiang Wang for their helpful discussions and comments. This work was supported by the Major State Basic Research Development Program in China under Grant No. 2014CB845400, and the National Natural Science Foundation of China under Grants No. 11522547, No. 11375251, and No. 11421505.
\end{acknowledgments}

\end{document}